# Model-Based Control for Power-to-X Platforms: Knowledge Integration for Digital Twins

Daniel Dittler
*Institute of Industrial Automation and Software Engineering*
University of Stuttgart
Stuttgart, Germany
daniel.dittler@ias.uni-stuttgart.de

Peter Frank
*Institute of Industrial Automation and Software Engineering*
University of Stuttgart
Stuttgart, Germany
peter.frank@ias.uni-stuttgart.de

Gary Hildebrandt
*Institute of Smart Systems and Services*
University of Pforzheim
Pforzheim, Germany
gary.hildebrandt@hs-pforzheim.de

Luisa Peterson
Dept. Process Systems Engineering
*MPI DCTS*
Magdeburg, Germany
peterson@mpi-magdeburg.mpg.de

Nasser Jazdi
*Institute of Industrial Automation and Software Engineering*
University of Stuttgart
Stuttgart, Germany
nasser.jazdi@ias.uni-stuttgart.de

Michael Weyrich
*Institute of Industrial Automation and Software Engineering*
University of Stuttgart
Stuttgart, Germany
michael.weyrich@ias.uni-stuttgart.de

*Abstract*— Offshore Power-to-X platforms enable flexible conversion of renewable energy, but place high demands on adaptive process control due to volatile operating conditions. To face this challenge, using Digital Twins in Power-to-X platforms is a promising approach. Comprehensive knowledge integration in Digital Twins requires the combination of heterogeneous models and a structured representation of model information. The proposed approach uses a standardized description of behavior models, semantic technologies and a graph-based model understanding to enable automatic adaption and selection of suitable models. It is implemented using a graph-based knowledge representation with Neo4j, automatic data extraction from Asset Administration Shells and port matching to ensure compatible model configurations.

## I. INTRODUCTION

Sustainable energy generation plays a central role in the reduction of $CO_2$ emissions and the transformation of the energy sector [1]. Offshore wind energy and Power-to-X technologies in particular offer great potential for a reliable and climate-friendly energy supply [2]. Offshore Power-to-X platforms in particular, which are operated without a direct grid connection and thus tap into remote locations, open up new business cases by converting renewable energy directly into storable energy carriers such as hydrogen or synthetic fuels [3]. However, offshore operation and the lack of a grid connection mean that the energy supply is directly dependent on volatile generation conditions, due to changing winds. This poses a particular challenge for dynamic process control, as fluctuations in the energy supply have a direct impact on the plant's operational management and require adaptive control. A promising approach to overcoming these challenges is model-based control [4]. By using physical, data-driven or hybrid models, different decision levels and time horizons can be covered - from short-term control to long-term optimization of plant operation [3]. However, the need to consider different models leads to increasing system complexity [5]. Not only do the models have to be considered separately, but their relations play a decisive role for a consistent and effective control strategy. In addition, the operating conditions change continuously due to the volatile environment and the models need to be adapted during the operating phase. This raises a key question: How can different behavior models be efficiently integrated and adapted during operation?

The paper therefore proposes an approach on how different models can be integrated in a model adaption approach to operate offshore Power-to-X platforms more robustly and efficiently.

The remainder of this contribution is structured as follows. Section II provides an overview of basics and related work. Section III describes the general overview, followed by the approach. Section IV describes the current implementation. Finally, a conclusion is drawn.

## II. BACKROUND AND RELATED WORK

This section first presents how different models at the various control levels of Power-to-X platforms may look like. The current state of research on Digital Twins, whose concept has the potential to enable the above-mentioned research question, is then presented. Possible approaches to integration are then analyzed and the research gap discussed.

### A. Models in the context of Power-to-X platforms

In process engineering, models can be categorized at different hierarchical levels, ranging from the molecular level to the production system level. These levels form a pyramid, with the molecular level at the base, followed by the phase level, process unit level, plant level, and finally the production system level at the top [6]. Each level represents a different degree of abstraction and serves specific purposes in modeling and simulation. For offshore Power-to-X platforms, this hierarchical structure is particularly relevant. At the production system level, flowsheet simulators provide a comprehensive representation of the entire Power-to-X platform, enabling the analysis of process chain configurations and their adaption to the offshore environment. These models are essential for evaluating the overall system performance and identifying optimal process operations with the necessary flexibility to handle the dynamic conditions of offshore environments (see, e.g., [7]). However, for integration into state-of-the-art rigorous optimization tools such as GAMS or Pyomo, more specialized models are required [8]. These models focus on individual process units, such as the Power-to-X reactor, and allow for detailed analysis and optimization of specific components. When the complexity of detailed models becomes a limiting factor, surrogate models offer a practical alternative. These models approximate the behavior of individual units or processes, either through simplified equations or purely data-driven approaches, enabling faster computations while maintaining sufficient accuracy for control and optimization purposes [4]. This hierarchical approach ensures that models are tailored to the specific requirements of different control levels, from high-level system optimization to detailed unit operation analysis.

### B. Digital Twins and current challenges

Digital Twins, as a virtual representation, encapsulate these behavior models as an essential part of their concept. As listed in [9], Digital Twins have so far mostly been realized monolithically. Especially when developing applications that interact with users or assets, the Digital Twin requires efficient and automated approaches to manage diverse models and data, without additional manual effort. These approaches highlight key research gaps, such as the need for intelligent algorithms for automating applications or for understanding different model implementations. As noted in [10], behavior models can provide significant value throughout the entire lifecycle of a Digital Twin. The authors also highlight the need for diverse forms of intelligence to enable the decomposition of monolithic system architectures. In this context, semantic technologies are proposed as promising solutions for modeling behavior aspects, as they can represent complex relationships between models. However, such semantic approaches have not yet been realized in Digital Twin implementations [11].

### C. Related work on meta-modelling of simulation models

This section looks at the state of the art for model integration or model provision. Only standards are considered, as these are the most promising for a comprehensive approach.

One standard is the Functional Mock-up Interface (FMI). The standard uses XML files, binary files and C code to provide a container functional mock-up (FMU). This encapsulates the models and provides variables via defined interfaces. The aim is to achieve simulation tool independence and to protect model-specific know-how. [12]

The Asset Administration Shell (AAS) makes it possible to provide models and data on various aspects of a system via standardized interfaces. This provision is specified by many sub-models with different information, such as technical data, manufacturer information or operating data. The sub-model simulation deals with the cross-company provision of simulation models. [13]

The Standard System Structure Parameterization (SSP) is a method independent of simulation tools for describing an overall system consisting of one or more components, e.g. individual FMUs together with their parameterization. SSP allows the definition of connections and hierarchies and thus a grouping into composite components by means of so-called system structure descriptions. It also defines how the parameterization data of the model should be saved and exchanged. The standard is based on a set of XML files that describe the component network including signal flow and parameterization. [14]

### D. Conclusion from backround and related work

Section II first described the various models used in Power-to-X platforms, thereby explaining the challenges presented in section I. The concept of the Digital Twin offers a fundamental solution for the provision of knowledge about model relations and for the execution of models. However, the current state of research shows that there is still considerable research potential for intelligent Digital Twins. Semantic technologies and automatic knowledge integration in particular offer promising approaches, but have rarely been implemented to date. Although there are standards for the provision of simulation knowledge, their meta-information overlaps, meaning that future consolidation should be examined. The AAS integrates meta-information on the hierarchy and on simulation models in two sub-models. Within these sub-models, reference could be made to SSP information and FMUs and tool-dependent models could also be described. However, a superordinate model understanding that considers aspects such as model categorization by level of detail or performance in different use cases has been lacking to date. In [16], the authors therefore presented an ontology for structuring a knowledge graph in order to enable automatic model adaption for Digital Twins. Automatic creation and adaption are required so that this knowledge graph does not have to be synchronized manually when the Digital Twin is reconfigured, which would result in considerable effort. Therefore, first an approach for the separation of knowledge provision is presented below, followed by the implementation of a method for automatic knowledge integration.

## III. APPROACH FOR AUTOMATED KNOWLEDGE INTEGRATION

In this section, the approach of automatic model adaption is presented first, then the focus of automatic knowledge provision is introduced. Fig. 1 shows the integration of model adaption into the classic control loop.

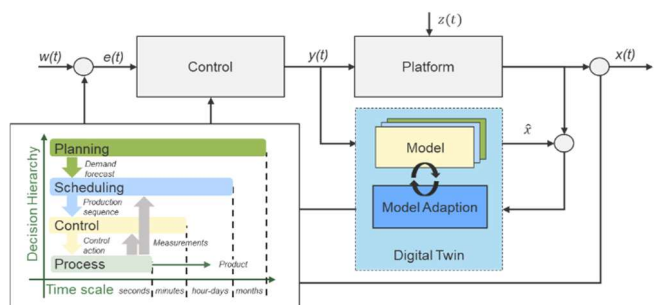

*Fig. 1 Digital Twin with Model Adaption for Decision Support [3]*

As outlined in the previous sections, model-based control plays an important role in the efficient control of Power-to-X platforms. Section II presents the variety of models at different decision levels (*Control*, *Scheduling*, *Planning*) in different tools with different computing times and accuracies. Depending on the change, different *models* can provide suitable solutions and optimize the operation of the platform. A *Model Adaption* is added to the control loop to provide the adapted controller parameters or new model configurations. Before applying the models in reality, the model adaption must first simulate them—both individually and in combination—to evaluate their suitabilities for the use case and prevent potential errors. To enable this comparison, the control variable $y(t)$ is transferred to the relevant models and the simulated behavior is compared with the actual behavior of the platform. In the event of a deviation $\hat{x}$, model adaption is triggered and models are parameterized or configured. The model adaption process leverages both the models and the Digital Twin's model comprehension to achieve this. The complexity and dynamics of Power-to-X processes, coupled with the large number of influencing variables and models, require a multi-level and networked approach to ensure a consistent way of working. In [15] the authors presented an approach for automatic model execution. The provision of the knowledge required for this has not yet been described. It is therefore necessary to link the models of a Power-to-X platform across several levels. The authors place the following requirements on the linking and the associated provision of knowledge in the Digital Twin:

- Standardized description of the models
- Structuring the heterogeneous model landscape
- Modeling of the model relations
- Scalable and expandable architecture

Fig. 2 shows the approach to knowledge provision in Digital Twins.

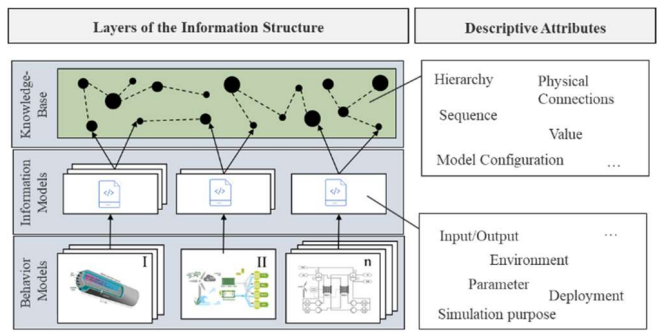

*Fig. 2 Structuring the information in Digital Twins [4], [7]*

At the *Information Models* level, the standardized description of the behavior models must be provided in machine-readable form, e.g. XML or JSON-based, to enable the automatic integration and read out for the model execution. Information such as the storage location for finding the models, solver properties, parameters and the simulation environment for execution must be described. The *Knowledge-Base* represents the model comprehension. For the superordinate model comprehension, as a partial aspect of the Digital Twin, the information models must include details such as the model's position within the plant topology, the classification of the discipline, the level of detail, and physical inputs and outputs. Only in this way is it possible to semantically link the different models in the model comprehension. In [16] the authors present an ontology-based approach for structuring this model comprehension, which also allows the extensibility of knowledge in the form of evaluation information of the different model configurations at runtime. The separation into the three layers, *Behavior Models*, *Information Models* and *Knowledge-Base*, ensures the necessary consistency, extensibility and scalability of the approach. In [15], the authors present the approach to automatic model execution that utilizes the knowledge provision presented here. In the following, the realization of the automatic knowledge integration is presented, which is necessary to enable the applicability of the approach.

## IV. REALIZATION

This section first introduces the realized model process and its application scenarios and then presents the method for automatic knowledge integration.

### A. Realization demonstrator and application scenario

A demonstrator was developed at the Institute for Industrial Automation and Software Engineering at the University of Stuttgart to evaluate the automation concepts for Power-to-X platforms. This consists of a modular design and allows the configuration of different process chains in the context of Power-to-X. Fig. 3 shows the demonstrator and the idea of realizing the AAS concept.

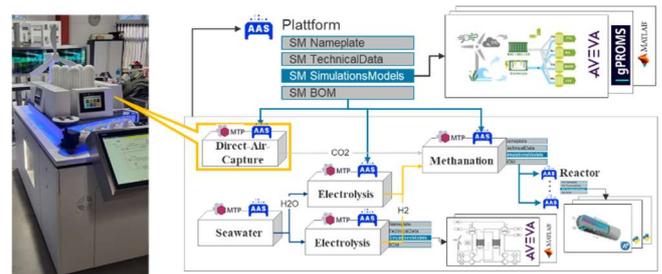

*Fig. 3 Realized demonstrator for future evaluation. [4], [7]*

For flexible configuration of the system configuration, the white containers in the figure can be exchanged on the platform. A module-type package approach is employed to enable the seamless integration of the process, while PCS Neo serves as the higher-level process control system. The higher-level control system is a S7-410, while the individual processes such as *Direct-Air-Capture*, *Electrolysis* and *Methanation* are controlled by CPU 1510SP installed in the containers. An IPC server is used as Hyper-V server for the process control system and the Digital Twin applications. The model process thus provides a real automation environment that allows the evaluation of use cases such as Plug & Produce, alarm management in teleoperation or Digital Twin approaches.

### B. Method for automatic knowledge integration

When the containers are replaced on the production platform, the models and relations in the Digital Twin must also be updated. Manual integrations are very time-consuming, which is why an automatic method is presented below. Fig. 4 shows this method and the tools used.

The automatic integration of knowledge takes place in three steps from left to right. The fourth step, the automatic model adaption that uses the knowledge, is described in more detail in [15].

The Asset Administration Shell is used to implement the information sources that represent the required information models. The AAS provides most of the required description features and also allows the integration of other standards such as FMI or SSP in the future. The Asset Administration Shells are hosted on an *AAS server*, which originates from the open-source project BaSyx. To process the Asset Administration Shells, the system reads AASX files, extracts relevant data and transfers it to a knowledge database structure. The *production sequence* is entered via a GUI depending on the system configuration and stored in a JSON. The relevant information is read from the AAS and the simulation sub-model and transferred to an object structure. The sub-model elements are identified by their ID-Short. After processing the individual Asset Administration Shells, the Bills of Material sub-model is read out and the hierarchy of assets is formed. The sub-model is modeled in *ArcheType* Full, so that the information of the hierarchy of the asset is described in the sub-model of the specific asset. It reads the sub-model and recursively builds the hierarchy which is added to the object structure.

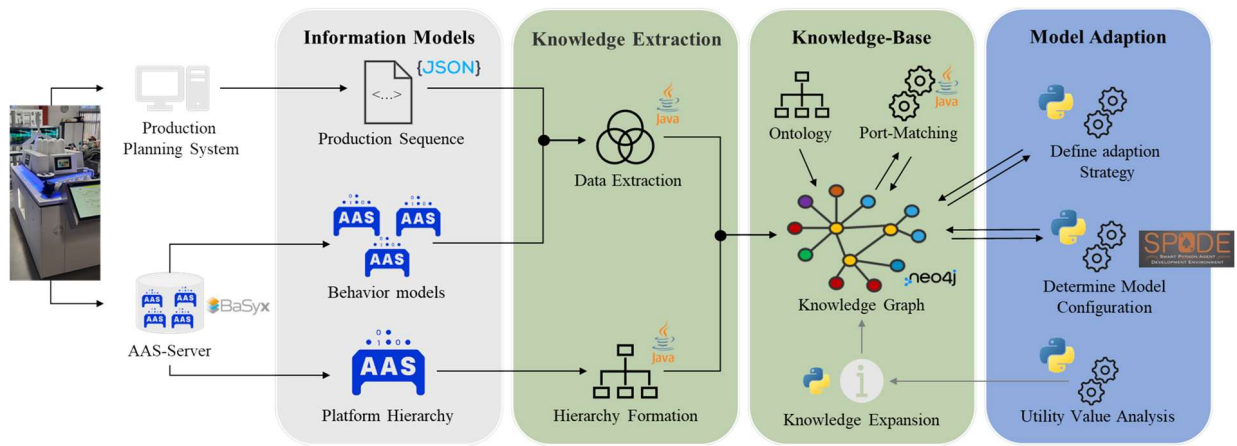

*Fig. 4 Realized method for the automatic integration of knowledge for model adaption in Digital Twins*

Neo4j was selected as the graph database for the implementation of model comprehension through a *Knowledge Graph*. Neo4j offers high flexibility, efficiency and extensive functionality. As a NoSQL database implemented in Java. Neo4j is characterized by its optimized architecture specifically designed for managing and traversing highly networked data. The ontology from [16] is used to structure the data. *Port-matching*, acting as a reasoner, extends the existing knowledge from the hierarchy and sequence with the links of physical ports. The hierarchy and sequence information serve as the initial constraint on the solution space. Models that have no topology connections to other models within a production system are excluded during the port-matching. Furthermore, a distinction is made between input ports and output ports. For all combinations of compatible ports then the units, range and data types are checked. Matching takes place if the units of the ports match or a conversion factor has been defined. The conversion factor is saved in the edge *connectsWith* between the two ports of the models. The standardized information sources allow deterministic knowledge extraction and thus provide automatic knowledge integration into a Digital Twin. The creation of the Asset Administration Shells is currently carried out by the simulation expert and represents the further potential for automation in this context. Large-Language-Model-based approaches for automatic generation of AAS may be applicable in future.

## V. CONCLUSION AND OUTLOOK

Offshore Power-to-X platforms enable sustainable energy conversion in remote locations but are highly dependent on volatile energy sources, requiring adaptive control. This article deals with:

- **Heterogeneous model landscape:** Various models at different hierarchy levels are essential for accurately representing and controlling.
- **Digital Twin**: Integrating these models into a Digital Twin ensures that they are connected and can be used adaptively.
- **Automated knowledge Integration:** Using Asset Administration Shell and a Knowledge Graph enables automated model adaption

This work highlights the benefits of automated model integration for adaptive control. Future work will focus on further embedding this approach into the demonstrator and evaluating its effectiveness.

ACKNOWLEDGMENT

This contribution was funded by the Federal Ministry of Research, Technology and Space (BMFTR) under grant 03HY302R.

REFERENCES

[1] D. Franzmann *et al.*, "Green hydrogen cost-potentials for global trade," *international journal of hydrogen energy*, vol. 48, no. 85, pp. 33062–33076, 2023.
[2] D. Dittler *et al.*, "Digitaler Zwilling für eine modulare Offshore-Plattform: Effizienzsteigerung grüner Power-to-X-Produktionsprozesse," *atp magazin*, 2022.
[3] P. Häbig *et al.*, "A Modular System Architecture for an Offshore Off-grid Platform for Climate-neutral Power-to-X Production in H2Mare," *Procedia CIRP*, 2024.
[4] L. Peterson, A. Forootani, E. I. Sanchez Medina, I. V. Gosea, K. Sundmacher, and P. Benner, "Towards Digital Twins for Power-to-X: Comparing Surrogate Models for a Catalytic CO2 Methanation Reactor," 2024.
[5] A. Ferrari and K. Willcox, "Digital twins in mechanical and aerospace engineering," *Nature Computational Science*, vol. 4, no. 3, pp. 178–183, 2024.
[6] H. Freund and K. Sundmacher, "Towards a methodology for the systematic analysis and design of efficient chemical processes: Part 1. From unit operations to elementary process functions," *Chemical Engineering and Processing: Process Intensification*, vol. 47, no. 12, pp. 2051–2060, 2008.
[7] P. Rentschler, C. Klahn, and R. Dittmeyer, "The Need for Dynamic Process Simulation: A Review of Offshore Power-to-X Systems," *Chemie Ingenieur Technik*, vol. 96, 1-2, pp. 114–125, 2024.
[8] N. Moessner, P. Haebig, and K. Hufendiek, "Developing a Modelling-Approach to represent Flexibility in Process Engineering‑Implementation of a Dynamic Scheduling for a Green Power-to-X Production," *Procedia CIRP*, vol. 126, pp. 508–512, 2024.
[9] G. Hildebrandt, D. Dittler, P. Habiger, R. Drath, and M. Weyrich, "Data integration for digital twins in industrial automation: A systematic literature review," *IEEE Access*, 2024.
[10] D. Dittler, V. Stegmaier, N. Jazdi, and M. Weyrich, "Illustrating the benefits of efficient creation and adaption of behavior models in intelligent Digital Twins over the machine life cycle," *Journal of Manufacturing Systems*, vol. 76, pp. 520–539, 2024.
[11] F. G. Listl, D. Dittler, G. Hildebrandt, V. Stegmaier, N. Jazdi, and M. Weyrich, "Knowledge graphs in the digital twin: A systematic literature review about the combination of semantic technologies and simulation in industrial automation," *IEEE Access*, 2024.
[12] Functional Mock-up Interface, *The leading standard to exchange dynamic simulation models*. [Online]. Available: https://fmi-standard.org/
[13] Industrial Digital Twin Assosiation, *Asset Administration Shell*. [Online]. Available: https://industrialdigitaltwin.org/
[14] SSP, *System Structure & Parameterization*. [Online]. Available: https://ssp-standard.org/
[15] D. Dittler, D. Stauss, P. Rentschler, J. Stümpfle, N. Jazdi, and M. Weyrich, "Flexible Co-Simulation Approach for Model Adaption in Digital Twins of Power-to-X Platforms," in *2024 IEEE 29th International Conference on Emerging Technologies and Factory Automation (ETFA)*, 2024, pp. 1–4.
[16] D. Dittler, F. Bodenstein, G. Hildebrandt, N. Jazdi, and M. Weyrich, "Automated Configuration of Behavior Models in Digital Twins based on a Knowledge-Graph," *Procedia CIRP*, vol. 130, pp. 683–688, 2024.